\documentstyle[12pt]{article}
\begin{document} \bibliographystyle{unsrt}
\vspace{3pc}
\begin{center}

{\Large \bf Lorentz Group in Condensed Matter
Physics}\footnote{Presented at the 4th International School on
Theoretical Physics: Symmetry and Structural Properties of Condensed
Matter (Zajaczkowo, Poland, August-September, 1996)}
\vspace{6mm}

Y.S.Kim\footnote{Internet: kim@umdhep.umd.edu}\\
{\it Department of Physics, University of Maryland,}\\
{\it College Park, Maryland 20742, U.S.A.}
\vspace{10mm}

\end{center}

\begin{abstract}

It is shown that the Lorentz group plays prominent roles in
at least two areas in condensed matter physics, namely in the Bogoliubov
transformation and optical filters.  It is pointed out that the
underlying symmetry of the Bogoliubov transformation is that of two
coupled oscillators, and that the underlying symmetry of two coupled
oscillators in that of the group $O(3,3)$.  The Lorentz group is also
shown to be the underlying symmetry group for the Jones matrix
formalism which is standard language for optical filters.
\end{abstract}

\section{Introduction}\label{intro}

The Lorentz group is definitely the language of Lorentz
transformations,\cite{wig39,barg47} and is therefore an indispensable
tool in high-energy particle physics which deals with particles moving
with speed very close to the speed of light.  I have built up my background
in the Lorentz group because I am a particle physicist.  However, this
group provides the underlying mathematics in many other branches of
physics.  Quantum optics is a case in point.  In connection with
two-photon coherent states, our colleagues in optics produced a new
scientific word ``squeezed states'' of
light.\cite{yuen76,cav85,knp91}

The Lorentz group is useful for squeezed states because Lorentz boosts
are squeeze transformations.\cite{dir49,yukawa53b}  Indeed, there are many
other squeeze transformations in physics, and the Lorentz group serves
as the hidden symmetry group in many branches of physics.  For instance,
linear canonical transformations in both quantum and classical mechanics
contain the symmetry of the Lorentz group.\cite{knp91}  While there
are many other possibilities, we should not overlook condensed matter
physics.

In this report, we shall discuss applications of the Lorentz group in
two distinct areas of condensed matter physics.  The first area is the
Bogoliubov transformation in superconductivity, and the other area is
the Jones matrix formalism governing symmetries of optical filters.
It is well known that the Bogoliubov transformation is a problem of
two coupled oscillators.  It is also known that the transformation is
like a Lorentz boost.  What is not well known is that the coupled
oscillator problem is a problems in the Lorentz group.

We can study much of crystal structures from optical properties of
matter.  Often we use them to construct optical filters.  The standard
mathematical tool for the filters is called the Jones matrix
formalism.\cite{jones41,swind75}
However, what is unknown is the fact that the Jones matrix formalism
is a representation of the Lorentz group.  In this note, we would like
to point out that the bilinear representation
of the six-parameter Lorentz group\cite{barg47} is the natural language
for polarization of light waves.

In Sec. \ref{o33}, we discuss the symmetry of two coupled
oscillators, which forms the basis for the Bogoliubov transformation.
It is pointed out that the symmetry is as rich as that of the Lorentz
group with three space-like dimensions and three time-like directions.
In Sec. \ref{lgpol}, it is shown that the two-by-two matrix formalism
of the Lorentz group is the natural language for optical filters.
This formalism reproduces the traditional mathematical device called
the Jones matrix formalism.

\section{Construction of the O(3,3) Symmetry Group from Two Coupled
Oscillators}\label{o33}
The Bogoliubov transformation is a well-established language in
condensed matter physics, and it does not seem to appear necessary
to explain what how this transformation works in the theory of
superconductivity.\cite{bogo58,fewa71,tink75}  On the other hand,
it is interesting to note that the transformation is basically the
problem of diagonalizing a system of two coupled oscillators.  Then
we can raise the following question.

The coupled-oscillator problem is discussed in Goldstein's textbook
on classical mechanics whose second edition was published in
1980.\cite{gold80}  Most of us have an impression that we can solve
the coupled oscillator problem using a rotation matrix, even though
Goldstein states in his book that the rotation alone is not enough.
It has been shown recently that Han {\it et al.} that it requires
the $O(3,3)$ symmetry in order to understand fully the coupled
oscillator problem.\cite{hkn95jm}   This section is based on the
work of Han {\it et al}.

We assume here that the reader is familiar with the basic fact that
one isolated oscillator has an $Sp(2)$ symmetry in phase space.
Thus, we should start with two independent $Sp(2)$ symmetries.  It
will be shown that an attempt to couple them by introducing a
rotation matrix leads to the $Sp(4)$ symmetry which Dirac discovered
in 1963.\cite{dir63}.  However, the $Sp(4)$ symmetry deals only with
canonical transformations.  If the size of the phase space is
enlarged or contracted, the transformation is not canonical.  In
this case, we need a large symmetry, and this symmetry group is
locally isomorphic to $O(3,3)$ or the Lorentz group with three
space-like and three time-like directions.

Let us consider a system of two coupled harmonic oscillators.  The
Hamiltonian for this system is
\begin{equation}\label{hamil}
H = {1\over 2}\left\{{1\over m}_{1}p^{2}_{1} +
{1\over m}_{2}p^{2}_{2} +
A x^{2}_{1} + B x^{2}_{2} + C x_{1} x_{2} \right\}.
\end{equation}
where
\begin{equation}
A > 0, \qquad B > 0, \qquad 4AB - C^2 > 0 .
\end{equation}

We are interested in transformations which will uncouple the above
Hamiltonian, or conversely those which will bring to the above
coupled form from two uncoupled one-oscillator Hamiltonian.
For the two
uncoupled oscillators, we can start with the coordinate system:
\begin{equation}\label{coord1}
\left(\eta _{1}, \eta _{2}, \eta _{3}, \eta _{4} \right) = \left(x_{1},
p_{1}, x_{2}, p_{2} \right) .
\end{equation}
This coordinate system is different from the traditional coordinate system
where the coordinate variables are ordered as $\left(x_{1}, x_{2}, p_{1},
p_{2} \right)$.  This unconventional coordinate system does not change the
physics or mathematics of the problem, but is convenient for studying
the uncoupled system as well as expanding and shrinking phase spaces.

Since the two oscillators are independent, it is possible to
perform linear canonical transformations on each coordinate separately.
The canonical transformation in the first coordinate system is generated by
\begin{equation}\label{eq.21}
A_{1} = {1 \over 2} \pmatrix{\sigma _{2} & 0 \cr 0 & 0 } , \qquad
B_{1} = {i \over 2} \pmatrix{\sigma _{3} & 0 \cr 0 & 0 } , \qquad
C_{1} = {i \over 2} \pmatrix{\sigma _{1} & 0 \cr 0 & 0 } .
\end{equation}
These generators satisfy the Lie algebra:
\begin{equation}\label{commuo21}
\left[A_{1}, B_{1}\right] = iC_{1}, \qquad
\left[B_{1}, C_{1}\right] = -iA_{1}, \qquad
\left[C_{1}, A_{1}\right] = iB_{1}. \qquad
\end{equation}
It is also well known that this set of commutation relations is
identical to that for the $(2 + 1)$-dimensional Lorentz group.	Linear
canonical transformations on the second coordinate are generated by
\begin{equation}\label{eq.23}
A_{2} = {1 \over 2} \pmatrix{0 & 0 \cr 0 & \sigma _{2}} , \qquad
B_{2} = {i \over 2} \pmatrix{0 & 0 \cr 0 & \sigma _{3}} , \qquad
C_{2} = {i \over 2} \pmatrix{0 & 0 \cr 0 & \sigma _{1}} .
\end{equation}
These generators also satisfy the Lie algebra of
Eq.(\ref{commuo21}).  We are interested here in constructing the symmetry
group for the coupled oscillators by soldering two $Sp(2)$ groups generated
by $A_{1}, B_{1}, C_{1}$ and $A_{2}, B_{2}, C_{2}$ respectively.

It will be more convenient to use the linear combinations:
\begin{eqnarray}
A_{+} &=& A_{1} + A_{2}, \qquad B_{+} = B_{1} + B_{2}, \qquad
C_{+} = C_{1} + C_{2}, \nonumber \\[2mm]
A_{-} &=& A_{1} - A_{2}, \qquad B_{-} = B_{1} - B_{2}, \qquad
C_{-} = C_{1} - C_{2},
\end{eqnarray}
These matrices take the form
\begin{eqnarray}
A_{+} &=& {1 \over 2}\pmatrix{\sigma _{2} & 0 \cr 0 & \sigma _{2}} , \qquad
A_{-} = {1 \over 2}\pmatrix{\sigma _{2} & 0 \cr 0 & -\sigma _{2}}, \nonumber \\[2mm]
B_{+} &=& {i \over 2}\pmatrix{\sigma _{3} & 0 \cr 0 & \sigma_{3}}, \qquad
B_{-} = {i \over 2}\pmatrix{\sigma _{3} & 0 \cr 0 & -\sigma _{3}}, \nonumber \\[2mm]
C_{+} &=& {i \over 2}\pmatrix{\sigma _{1} & 0 \cr 0 & \sigma _{1}}, \qquad
C_{-} = {i \over 2}\pmatrix{\sigma _{1} & 0 \cr 0 & -\sigma _{1}}.
\end{eqnarray}
The sets $\left(A_{+}, B_{+}, C_{+}\right)$ and
$\left(A_{+}, B_{-}, C_{-}\right)$ satisfy the Lie algebra of
Eq.(\ref{commuo21}).  The same is true for $\left(A_{-}, B_{+}, C_{-}\right)$
and $\left(A_{-}, B_{-}, C_{+}\right)$.

Next, let us couple the oscillators through a rotation generated by
\begin{equation}\label{A0}
A_{0} = {i \over 2}\pmatrix{0 & -I \cr I & 0} .
\end{equation}
In view of the fact that the first two coordinate variables are for the phase
space of the first oscillator, and the third and fourth are for the second
oscillator, this matrix generates parallel rotations in the $(x_{1},x_{2})$
and $(p_{1},p_{2})$ coordinates.  As the coordinates $(x_{1},x_{2})$ are
coupled through a two-by-two matrix, the coordinate $(p_{1},p_{2})$ are
coupled through the same two-by-two matrix.

Then, $A_{0}$ commutes with $A_{+}, B_{+}, C_{+}$, and the following
commutation relations generate new operators $A_{3}, B_{3}$ and
$C_{3}$:
\begin{equation}\label{eq.27}
\left[A_{0}, A_{-}\right] = iA_{3}, \qquad
\left[A_{0}, B_{-}\right] = iB_{3}, \qquad
\left[A_{0}, C_{-}\right] = iC_{3}, \qquad
\end{equation}
where
\begin{equation}\label{eq.28}
A_{3} = {1 \over 2}\pmatrix{0 & \sigma _{2} \cr \sigma _{2} & 0} , \qquad
B_{3} = {i \over 2}\pmatrix{0 & \sigma _{3} \cr \sigma _{3} & 0} , \qquad
C_{3} = {i \over 2}\pmatrix{0 & \sigma _{1} \cr \sigma _{1} & 0} .
\end{equation}

In this section, we started with the generators of the symmetry groups for
two independent oscillators.  They are $A_{1}, B_{1}, C_{1}$ and
$A_{2}, B_{2}, C_{2}$.	We then introduced $A_{0}$ which generates coupling
of two oscillators.  This processes produced three additional generators
$A_{3}, B_{3}, C_{3}$.	It is remarkable that $C_{3}, B_{3}$ and $A_{+}$
form the set of generators for another $Sp(2)$ group.  They satisfy the
commutation relations
\begin{equation}
[B_{3}, C_{3}] = -iA_{+} , \qquad  [C_{3}, A_{+}] = iB_{3} , \qquad
[A_{+}, B_{3}] = iC_{3} .
\end{equation}
The same can be said about the sets $A_{+}, B_{1}, C_{1}$ and
$A_{+}, B_{2}, C_{2}$.	These Sp(2)-like groups are associated with the
coupling of the two oscillators.

For a dynamical system consisting of two pairs of canonical variables
$x_{1}, p_{1}$ and $x_{2}, p_{2}$, we have introduced the coordinate
system $\left(\eta _{1}, \eta _{2}, \eta _{3}, \eta _{4} \right)$
defined in Eq.(\ref{coord1}).  The transformation of the variables
from $\eta _{i}$ to $\xi _{i}$ is canonical if
\begin{equation}\label{symp}
M J \tilde{M} = J ,
\end{equation}
where
$$
M_{ij} = {\partial \over \partial \eta _{j}}\xi _{i},
$$
\noindent and
\begin{equation}\label{JKQ}
J = \pmatrix{0 & 1 & 0 & 0 \cr -1 & 0 & 0 & 0
\cr 0 & 0 & 0 & 1 \cr 0 & 0 & -1 & 0} .
\end{equation}
This form of the J matrix appears different from the traditional literature,
because we are using the new coordinate system.  In order to avoid possible
confusion and to maintain continuity with our earlier publications, we give
in the Appendix the expressions for the J matrix and the ten generators of
the $Sp(4)$ group in the traditional coordinate system.  There are four
rotation generators and six squeeze generators in this group.

In this new coordinate system, the rotation generators take the form
$$
L_{1} = {-1\over 2}\pmatrix{0&\sigma _{2} \cr \sigma _{2} & 0}, \qquad
L_{2} = {i\over 2}\pmatrix{ 0 & -I \cr I & 0}  ,
$$
\begin{equation}\label{rotKQ}
L_{3} = {-1\over 2}\pmatrix{\sigma _{2}&0 \cr 0 &-\sigma _{2}}, \qquad
S_{3} = {1\over 2}\pmatrix{\sigma _{2} & 0 \cr 0 & \sigma _{2} } .
\end{equation}
The squeeze generators become
$$
K_{1} = {i\over 2}\pmatrix{\sigma _{1} & 0 \cr 0 & -\sigma _{1} },
\quad
K_{2} = {i\over 2}\pmatrix{\sigma _{3} & 0 \cr 0 & \sigma _{3}} , \quad
K_{3} = -{i\over 2}\pmatrix{0 & \sigma _{1} \cr \sigma _{1} & 0} ,
$$

\begin{equation}\label{sqKQ}
Q_{1} = {i\over 2}\pmatrix{-\sigma _{3} & 0 \cr 0 & \sigma _{3}}, \quad
Q_{2} = {i\over 2}\pmatrix{\sigma _{1} & 0 \cr 0 & \sigma _{1}} , \quad
Q_{3} = {i\over 2}\pmatrix{0 & \sigma _{3} \cr \sigma _{3} & 0} .
\end{equation}

There are now ten generators.  They form the Lie algebra for the $Sp(4)$
group:
$$
[L_{i}, L_{j}] = i\epsilon _{ijk} L_{k}, \qquad
[L_{i}, S_{3}] = 0, \qquad
$$
$$
[L_{i}, K_{j}] = i\epsilon _{ijk} K_{k}, \qquad
[L_{i}, Q_{j}] = i\epsilon _{ijk} Q_{k}, \qquad
$$
$$
[K_{i}, K_{j}] = [Q_{i}, Q_{j}] = -i\epsilon _{ijk} L_{k} , \qquad
[K_{i}, Q_{j}] = -i\delta _{ij} S_{3} ,
$$
\begin{equation}\label{LieSp4}
[K_{i}, S_{3}] = -iQ_{i} , \qquad [Q_{i}, S_{3}] = iK_{i} .
\end{equation}

Indeed, these matrices can be identified with the $A, B$, and $C$
matrices derived from the coupled oscillators in the following manner.
$$
A_{+} = S_{3}, \qquad A_{-} = -L_{3}, \qquad A_{3} = -L_{1} ,  \qquad
A_{0} = L_{2},
$$
$$
B_{+} = K_{2}, \qquad B_{-} = -Q_{1}, \qquad B_{3} = Q_{3} , \\
$$
\begin{equation}
C_{+} = Q_{2}, \qquad C_{-} = K_{1}, \qquad C_{3} = -K_{3} .
\end{equation}

In this section, we started with the $Sp(2)$ symmetry for each of the
two oscillator,
and introduced the parallel rotation to couple the system.  It is
interesting to note that this process leads to the $Sp(4)$ symmetry.

The $A_{0}$ matrix given in Eq.(\ref{A0}) generates the coupling of
two phase spaces by rotation.  Within this coordinate system, we are
interested in relative adjustments of the sizes of the two phase
spaces.  By making this adjustment, we are changing the relative
size of the two phase spaces.  This is not a canonical transformation,
but is quite relevant to the physics with dissipation or with input
energy from external sources.

For this purpose, we need the generators of the form
\begin{equation}\label{G3}
G_{3} = {i\over 2} \pmatrix{I & 0 \cr 0 & -I} .
\end{equation}
This matrix generates scale transformations in phase space.  The
transformation leads to a radial expansion of the phase space of the
first coordinate\cite{kili89} and contracts the phase space of the
second coordinate.  What is the physical significance of this
operation?  Classically, it is a routine procedure.  In quantum
mechanics, the expansion of phase space leads to an increase in
uncertainty and entropy.  Mathematically speaking, the contraction of
the second coordinate should cause a decrease in uncertainty and entropy.
Can this happen?  The answer is clearly No, because it will violate the
uncertainty principle.	This problem requires further research.

In the meantime, let us study what happens when the matrix $G_{3}$ is
introduced into the set of matrices given in Eq.(\ref{rotKQ}) and
Eq.(\ref{sqKQ}).  It commutes with $S_{3}, L_{3}, K_{1}, K_{2}, Q_{1}$,
and $Q_{2}$.  However, its commutators with the rest of the matrices produce
four more generators:
\begin{eqnarray}
\left[G_{3}, L_{1}\right] &=& iG_{2} , \qquad
\left[G_{3}, L_{2}\right] = -iG_{1} ,\nonumber \\[2mm]
\left[G_{3}, K_{3}\right] &=& iS_{2} , \qquad
\left[G_{3}, Q_{3}\right] = -iS_{1} ,
\end{eqnarray}
with
\begin{eqnarray}
G_{1} &=& {i\over 2}\pmatrix{0 & I \cr I & 0} , \qquad
G_{2} = {1\over 2}\pmatrix{0 & -\sigma_{2} \cr \sigma_{2} & 0} , \nonumber \\[2mm]
S_{1} &=& {-i\over 2}\pmatrix{0 & -\sigma_{3} \cr \sigma_{3} & 0} , \qquad
S_{2} = {i\over 2}\pmatrix{0 & -\sigma_{1} \cr \sigma_{1} & 0} .
\end{eqnarray}
If we take into account the above five generators in addition to the
ten generators of $Sp(4)$, there are fifteen generators.  These generators
satisfy the following set of commutation relations.
$$
[L_{i}, L_{j}] = i\epsilon _{ijk} L_{k}, \qquad
[S_{i}, S_{j}] = i\epsilon _{ijk} S_{k}, \qquad
[L_{i}, S_{j}] = 0, \qquad
$$
$$
[L_{i}, K_{j}] = i\epsilon _{ijk} K_{k}, \qquad
[L_{i}, Q_{j}] = i\epsilon _{ijk} Q_{k}, \qquad
[L_{i}, G_{j}] = i\epsilon _{ijk} G_{k},
$$
$$
[K_{i}, K_{j}] = [Q_{i}, Q_{j}] = [Q_{i}, Q_{j}]
= -i\epsilon _{ijk} L_{k} , \qquad
$$
$$
[K_{i}, Q_{j}] = -i\delta _{ij} S_{3} , \qquad
[Q_{i}, G_{j}] = -i\delta _{ij} S_{1} , \qquad
[G_{i}, K_{j}] = -i\delta _{ij} S_{2} .
$$
$$
[K_{i}, S_{3}] = -iQ_{i} , \qquad [Q_{i}, S_{3}] = iK_{i} ,\qquad
[G_{i}, S_{3}] = 0 ,
$$
$$
[K_{i}, S_{1}] = 0 , \qquad [Q_{i}, S_{1}] = -iG_{i} ,\qquad
[G_{i}, S_{1}] = iQ_{i} ,
$$

\begin{equation}
[K_{i}, S_{2}] = iG_{i} , \qquad [Q_{i}, S_{2}] = 0 ,\qquad
[G_{i}, S_{2}] = -iK_{i} .
\end{equation}

Indeed, the ten $Sp(4)$ generators together with the five new generators
form the Lie algebra for the group $SL(4,r)$.  This group is known to
be locally isomorphic to the Lorentz group $O(3,3)$ with three space
variables and three time variables.  Indeed, we can study the coupled
oscillator problem in terms of the Lorentz group.  Conversely, we can
use the oscillator problem in order to understand the Lorentz group.
It would be a challenging problem to see whether there is a larger
symmetry in superconductivity than what is known today.

In this connection, we should note that there have been many papers
in recent on squeezed states of light which deal with coupling of
two-particle systems, and the physics of squeezed states is strikingly
similar to the physics of superconductivity.\cite{bishop88,vourdas92}
As in the case of squeezed states, the basic symmetry of the Bogoliubov
transformation is that of the $Sp(4)$ group.\cite{kibir88}  We should
study more along this direction.

\section{Lorentz Group for Optical Filters}{\label{lgpol}}
There is another area of condensed matter physics where the Lorentz
group serves as the basic language.  Anisotropic optical filters
are due to crystal structures.  Indeed for optical filters, a
systematic mathematics has been developed.\cite{swind75}  Since Jones
was the one who initiated this process,\cite{jones41} the formalism
is called the Jones matrix formalism.  There also have been attempts
to show that this formalism constitutes a representation of the Lorentz
group.\cite{parent60}

Recently, Han {\it et al.}\cite{hkn96} used the algebra of squeezed
states\cite{knp91} to reproduce the Jones matrix formalism.  The
algebra Han {\it et al.} used in their paper is a two-by-two
representation of the six-parameter Lorentz group.  They did not
realize that what they did was the Jones matrix formalism, but they
got all the ingredient to reach this conclusion in their paper.
They even went further to show that the bilinear representation
of the Lorentz group is the most appropriate language for the
Jones matrix formalism.  In this section, we shall review the work
done by Han {\it et al}.

In studying polarized light propagating along the $z$ direction,
the traditional approach is to consider the $x$
and $y$ components of the electric fields.  Their amplitude ratio and the
phase difference determine the degree of polarization.  Thus, we can
change the polarization either by adjusting the amplitudes, by changing
the relative phase shift, or both.  For convenience, we call the optical
device which changes amplitudes an ``attenuator,'' and the device which
changes the relative phase a ``phase shifter.''

Let us write the electric field vector as
\begin{eqnarray}\label{cosine}
E_{x} &=& A \cos{\left(kz - \omega t + \phi_{1}\right)} , \cr
E_{y} &=& B \cos{\left(kz - \omega t + \phi_{2}\right)} ,
\end{eqnarray}
where $A$ and $B$ are the amplitudes which are real and positive
numbers, and $\phi_{1}$ and $\phi_{2}$ are the phases of the $x$ and
$y$ components respectively.  This form is useful not only in
classical optics but also applicable to coherent and squeezed states
of light.\cite{knp91}

The traditional language for this two-component light is the Jones
matrix formalism which is discussed in standard optics
textbooks.\cite{predo93} In this formalism, the above two components
are combined into one column matrix with the exponential form for the
sinusoidal function.
\begin{equation}\label{expo1}
\pmatrix{E_{x} \cr E_{y}} =
\pmatrix{A \exp{\left\{i(kz - \omega t + \phi_{1})\right\}}  \cr
B \exp{\left\{i(kz - \omega t + \phi_{2})\right\}}} .
\end{equation}
The content of polarization is determined by the ratio:
\begin{equation}
{E_{y}\over E_{x}} = \left({B\over A}\right) e^{i(\phi_{2} - \phi_{1})} .
\end{equation}
which can be written as one complex number:
\begin{equation}\label{ratio}
w = r e^{i\phi}
\end{equation}
with
$$
r = {B \over A} , \qquad \phi = \phi_{2} - \phi_{1} .
$$
The degree of polarization is measured by these two real numbers, which are
the amplitude ratio and the phase difference respectively.

The purpose of this paper is to discuss the transformation properties of
this complex number $w$.  The transformation takes place when the light
beam goes through an optical filter whose transmission properties are not
isotropic.  There are two transverse directions which are perpendicular to
each other.  The absorption coefficient in one transverse direction could
be different from the coefficient along the other direction.  Thus, there
is the ``polarization'' coordinate in which the absorption can be described
by
\begin{equation}\label{atten}
\pmatrix{e^{-\eta_{1}} & 0 \cr 0 & e^{-\eta_{2}}} =
e^{-(\eta_{1} + \eta_{2})/2} \pmatrix{e^{\eta/2} & 0 \cr 0 & e^{-\eta/2}}
\end{equation}
with $\eta = \eta_{2} - \eta_{1}$ .
This attenuation matrix tells us that the electric fields are attenuated at
two different rates.  The exponential factor $e^{-(\eta_{1} + \eta_{2})/2}$
reduces both components at the same rate and does not affect the degree of
polarization.  The effect of polarization is solely determined by the
squeeze matrix
\begin{equation}\label{sq1}
S(0, \eta) = \pmatrix{e^{\eta/2} & 0 \cr 0 & e^{-\eta/2}} .
\end{equation}
This type of mathematical operation is quite familiar to us from squeezed
states of light, if not from Lorentz boosts of spinors.  For convenience,
we call the above matrix an attenuator.

If the polarization coordinate is the same as the $xy$ coordinate where
the electric field components take the form of Eq.(\ref{cosine}), the
above attenuator is directly applicable to the column matrix of
Eq.(\ref{expo1}).  If the polarization coordinate is rotated by an angle
$\theta/2$, or by the matrix
\begin{equation}
R(\theta) = \pmatrix{\cos(\theta/2) & -\sin(\theta/2)
\cr \sin(\theta/2) & \cos(\theta/2)} ,
\end{equation}
then the squeeze matrix becomes $S(\theta, \eta) =
R(\theta) S(0, \eta) R(-\theta)$, where
\begin{equation}\label{sq2}
\pmatrix{e^{\eta/2}\cos^{2}(\theta/2) + e^{-\eta/2}\sin^{2}(\theta/2) &
(e^{\eta/2} - e^{-\eta/2})\cos(\theta/2) \sin(\theta/2) \cr
(e^{\eta/2} - e^{-\eta/2})\cos(\theta/2) \sin(\theta/2)
& e^{-\eta/2}\cos^{2}(\theta/2) + e^{\eta/2}\sin^{2}(\theta/2)} .
\end{equation}

Another basic element is the optical filter with two different values
of the index of refraction along the two orthogonal directions.  The
effect of this filter can be written as
\begin{equation}\label{phase}
\pmatrix{e^{i\lambda_{1}} & 0 \cr 0 & e^{i\lambda_{2}}}
= e^{-i(\lambda_{1} + \lambda_{2})/2}
\pmatrix{e^{-i\lambda/2} & 0 \cr 0 & e^{i\lambda/2}} ,
\end{equation}
with $\lambda = \lambda_{2} - \lambda_{1}$ .
In measurement processes, the overall phase factor
$e^{-i(\lambda_{1} + \lambda_{2})/2}$
cannot be detected, and can therefore be deleted.  The polarization
effect of the filter is solely determined by the matrix
\begin{equation}\label{shif1}
P(0, \lambda) = \pmatrix{e^{-i\lambda/2} & 0 \cr 0 & e^{i\lambda/2}} .
\end{equation}
This phase-shifter matrix appears like a rotation matrix around the
$z$ axis in the theory of rotation groups, but it plays a different
role in this paper.  We shall hereafter call this matrix a phase shifter.

Here also, if the polarization coordinate makes an angle $\theta$ with
the $xy$ coordinate system, the phase shifter becomes
$P(\theta, \lambda) = R(\theta) P(0, \lambda) R(-\theta)$ which takes
the form
\begin{equation}\label{shif2}
\pmatrix{e^{-i\lambda/2}\cos^{2}(\theta/2) +
e^{i\lambda/2}\sin^{2}(\theta/2) &
(e^{-i\lambda/2} - e^{i\lambda/2})\cos(\theta/2) \sin(\theta/2) \cr
(e^{-i\lambda/2} - e^{i\lambda/2})\cos(\theta/2) \sin(\theta/2)
& e^{i\lambda/2}\cos^{2}(\theta/2) + e^{-i\lambda/2}\sin^{2}(\theta/2)} .
\end{equation}

Since we are interested in repeated applications of these two different
kinds of matrices with different parameters, we shall work with the
generators of these transformations.  Let us introduce the Pauli spin
matrices of the form
\begin{equation}
\sigma_{1} = \pmatrix{1 & 0 \cr 0 & -1} , \quad
\sigma_{2} = \pmatrix{0 & 1 \cr 1 & 0} , \quad
\sigma_{3} = \pmatrix{0 & -i \cr i & 0} .
\end{equation}
These matrices are written in a different convention.  Here  $\sigma_{3}$
is imaginary, while $\sigma_{2}$ is imaginary in the traditional notation.
Also in this convention, we can construct three rotation generators
\begin{equation}
J_{i} = {1 \over 2} \sigma_{i} ,
\end{equation}
which satisfy the closed set of commutation relations
\begin{equation}\label{comm1}
\left[J_{i}, J_{j}\right] = i \epsilon_{ijk} J_{k} .
\end{equation}
We can also construct three boost generators
\begin{equation}
K_{i} = {i \over 2} \sigma_{i} ,
\end{equation}
which satisfy the commutation relations
\begin{equation}\label{comm2}
\left[K_{i}, K_{j}\right] = -i \epsilon_{ijk} J_{k} .
\end{equation}
The $K_{i}$ matrices alone do not form a closed set of commutation
relations, and the rotation generators $J_{i}$ are needed to form a
closed set:
\begin{equation}\label{comm3}
\left[J_{i}, K_{j}\right] = i \epsilon_{ijk} K_{k} .
\end{equation}

The six matrices $J_{i}$ and $K_{i}$ form a closed set of commutation
relations, and they are like the generators of the Lorentz group applicable
to the (3 + 1)-dimensional Minkowski space.  The group generated by the
above six matrices is called $SL(2,c)$ consisting of all two-by-two complex
matrices with unit determinant.

If we consider only the phase shifters, the mathematics is
basically repeated applications of $J_{1}$ and $J_{2}$, resulting in
applications also of $J_{3}$.  Thus, the phase-shift filters form an
$SU(2)$ or $O(3)$-like subgroup of the group $SL(2,c)$.  On the other
hand, if we consider only the attenuators, the mathematics
consists of repeated applications of $K_{1}$ and $J_{3}$, resulting in
applications also of $K_{2}$.  This is evident from the commutation
relation
\begin{equation}
\left[J_{3}, K_{1}\right] = i K_{2} .
\end{equation}
Indeed, $J_{3}, K_{1}$ and $K_{2}$ form a closed set of commutation
relations for the $Sp(2)$ or $O(2,1)$-like subgroup of
$SL(2,c)$.\cite{kitano89}.  This three-parameter subgroup has been
extensively discussed in connection with squeezed states of
light.\cite{yuen76,knp91}

If we use both the attenuators and phase shifters, the result is the
full $SL(2,c)$ group with six parameters.  The transformation matrix is
usually written as
\begin{equation}\label{abgd1}
L = \pmatrix{\alpha & \beta \cr \gamma & \delta} ,
\end{equation}
with the condition that its determinant be one:
$\alpha\delta - \gamma\beta$ = 1.  The repeated application of
two matrices of this kind results in
\begin{equation}\label{abgd2}
\pmatrix{\alpha_{2} & \beta_{2} \cr \gamma_{2} & \delta_{2}}
\pmatrix{\alpha_{1} & \beta_{1} \cr \gamma_{1} & \delta_{1}}
= \pmatrix{\alpha_{2}\alpha_{1} + \beta_{2}\gamma_{1} &
\alpha_{2}\beta_{1} + \beta_{2}\delta_{1} \cr
\gamma_{2}\alpha_{1} + \delta_{2}\gamma_{1} &
\gamma_{2}\beta_{1} + \delta_{2}\delta_{1}} .
\end{equation}
The most general form of the polarization transformation is the application
of this algebra to the column matrix of Eq.(\ref{expo1}).

We can obtain the same algebraic result by using the bilinear
transformation:
\begin{equation}\label{bilin1}
w' = {\delta w + \gamma \over \beta w + \alpha} .
\end{equation}
The repeated applications of these two transformations can be achieved
from
\begin{equation}\label{bilin2}
w_{1} = {\delta_{1} w + \gamma_{1} \over \beta_{1} w + \alpha_{1}} , \quad
w_{2} = {\delta_{2} w_{1} + \gamma_{2} \over \beta_{2} w_{1} + \alpha_{2}} .
\end{equation}
Then, it is possible to write $w_{2}$ as a function of $w$, and the result
is
\begin{equation}\label{bilin3}
w_{2} = {(\gamma_{2}\beta_{1} + \delta_{2}\delta_{1} ) w +
(\gamma_{2}\alpha_{1} + \delta_{2}\gamma_{1})
\over (\alpha_{2}\beta_{1} + \beta_{2}\delta_{1}) w +
(\alpha_{2}\alpha_{1} + \beta_{2}\gamma_{1})} .
\end{equation}
This is a reproduction of the algebra given in the matrix multiplication
of Eq.(\ref{abgd2}).  The form given in Eq.(\ref{bilin1}) is the bilinear
representation of the Lorentz group.\cite{barg47}

Let us go back to physics.  If we apply the matrix $L$ of Eq.(\ref{abgd1})
to the column vector of Eq.(\ref{expo1}), then
\begin{equation}
\pmatrix{\alpha & \beta \cr \gamma & \delta} \pmatrix{E_{x} \cr E_{y}}
= \pmatrix{\alpha E_{x} + \beta E_{y} \cr \gamma E_{x} + \delta E_{y}} ,
\end{equation}
which gives
\begin{equation}
{E'_{y} \over E'_{x}} = {\gamma E_{x} + \delta E_{y} \over
\alpha E_{x} + \beta E_{y}} .
\end{equation}
In term of the physical quantity $w$ defined in Eq.(\ref{ratio}), this
formula becomes
\begin{equation}
w' = {\gamma + \delta w \over \alpha + \beta w} .
\end{equation}
This equation is identical to the bilinear form given in
Eq.(\ref{bilin1}), and the ratio $w$ can now be identified with the
$w$ variable defined as the parameter of the bilinear representation
of the Lorentz group in the same equation.  As we stated before,
the purpose of this paper was to derive the transformation property
of this complex number, and the purpose has now been achieved.
The bilinear representation of the Lorentz group is clearly the
language of optical filters and resulting polarizations.

Here we restricted ourselves to the cases where the polarization axes
are are orthogonal and the polarization planes is perpendicular to
the direction of propagation.  But this is not always true, and the
physics becomes more complicated if more complex crystals are
taken into consideration.  If we combine the group theory of
polarization and the group theory of crystals, the result will be
a group extension.  This will introduce many new symmetry problems
to condensed matter physics.

\end{document}